# First experimental research of low energy proton radiography

Wei Tao(魏涛)[1)]    Yang Guojun(杨国君)    Long Jidong(龙继东)
He Xiaozhong(何小中)    Li Yiding(李一丁)    Zhang Xiaoding(张小丁)
Ma Chaofan(马超凡)    Zhao Liangchao(赵良超)    Shi Jinshui(石金水)

(*Institute of Fluid physics, CAEP, P.O. Box 919-106, Mianyang 621900, China*)

**Abstract:** Proton radiography is a new scatheless diagnostic tool, and which provides a potential development direction for advanced hydrotesting. Recently a low energy proton radiography system has been developed at CAEP. This system has been designed to use 11MeV proton beam to radiograph thin static objects. This system consists of a proton cyclotron coupled to an imaging beamline. The design features and commissioning results of this radiography system are presented.

## 1、Introduction

The technique of proton radiography (pRad) has been developed at LANL to utilize 800 MeV proton beam as a radiographic probe for the diagnosis of hydrotesting research [1]. The beam optics requirements for pRad lens system are well understood [2] and schematically shown in Figure 1, the different colors represent the different scattering angles within the object.

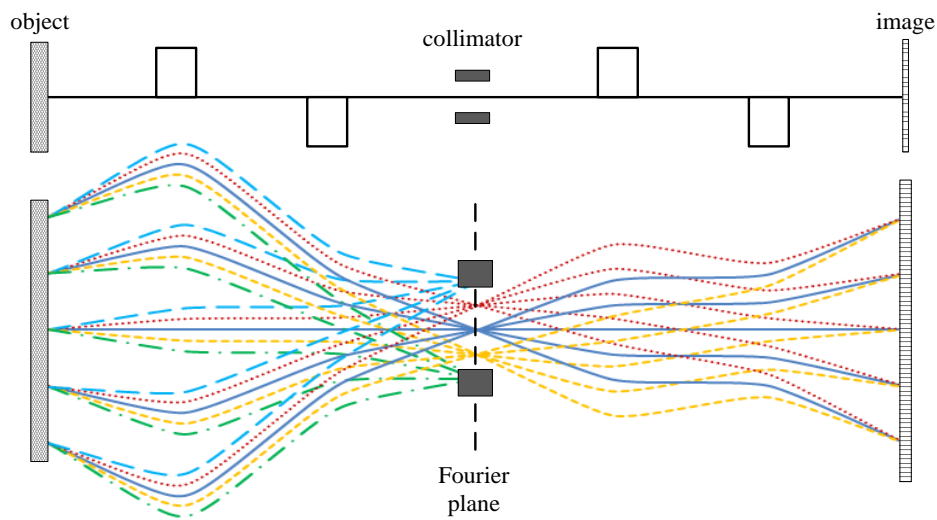

Fig.1    schematically diagram of a pRad lens system showing the point-to-point focus of particle trajectories (colored lines) from object to image.

There are two primary requirements of the pRad lens system [3]. First, the lens system must provide a point-to-point focus from object to image. Second, it must form a Fourier plane, where particles can be sorted by the scattering angle. With the latter requirement, particles with large scattering angles can be removed through transverse collimation at the Fourier plane. Otherwise, the object length must be suited with the beam energy, and the aperture of the lens system must be chosen to provide sufficient acceptance throughout the required field-of-view.

*supported by National Natural Science Foundation of China (11205144) and National Natural Science Foundation of China (11176001).
1) E-mail: weitaocaep@gmail.com

## 2、Low energy proton radiography system

Currently the pRad technique is mainly used in high energy proton accelerators, for example 800MeV pRad at LANSCE, 24GeV pRad at AGS [4] and 70GeV pRad at IHEP [5]. Recent work at Chinese Academy of Engineering Physics has extended this diagnostic technique to low energy proton accelerator.

### 2.1　Proton source

An 11MeV $H^-$ compact cyclotron used for medical radioactive isotope production is under operation [6] at Institute of Fluid Physics, CAEP. This cyclotron is capable of providing continuous wave beam, and the average current is about 50μA. The negative hydrogen beam is converted to protons through carbon stripping foil.

### 2.2　pRad imaging beamline

As shown in figure 2 is the layout of 11MeV low energy pRad imaging beamline, and figure 3 shows the twiss parameters. Firstly, proton beam passes through the matching section, and the beam is modulated for the correlation between particles' transverse displacement and angle deviation [3]. Then the modulated beam penetrate thin object while keeping the MCS angle and energy loss small enough to allow good spatial resolution. Finally, the beam passes through Zumbro lens system and arrives scintillator detector.

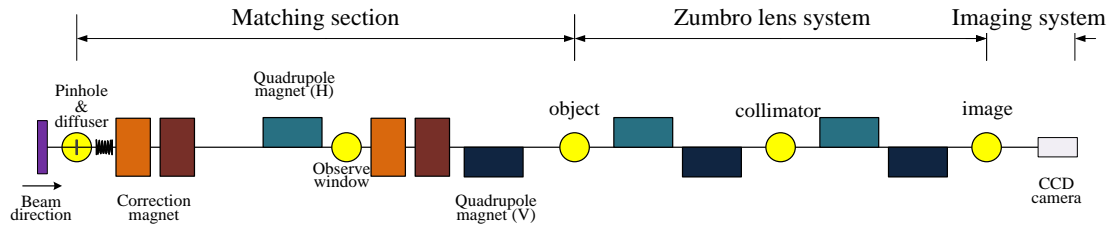

Fig.2　Layout of 11MeV low energy pRad beamline

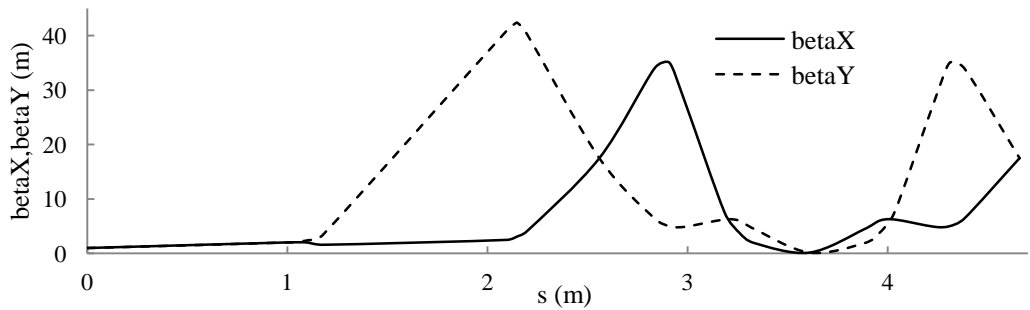

Fig.3　Twiss parameters of 11MeV low energy pRad beamline

The required phase space correlations are provided by a matching section just upstream of the object. Moreover, this section must also expand the beam's transverse size to fully illuminate the field-of-view. The proton beam from cyclotron passes through a pinhole with 1mm diameter first, the beam is truncated and the central protons are preserved. The preserved proton beam passes through a 20μm aluminum foil diffuser, which gives a small angular divergence to the beam and then passes through a set of magnets, which introduces a correlation between the radial position of proton in the object plane and its angle. What was shown in figure 4 is the evolution diagram of low energy proton beam in the matching section. Through the matching section, the proton beam is well modulated and the width is enlarged long enough to illuminate the whole object.

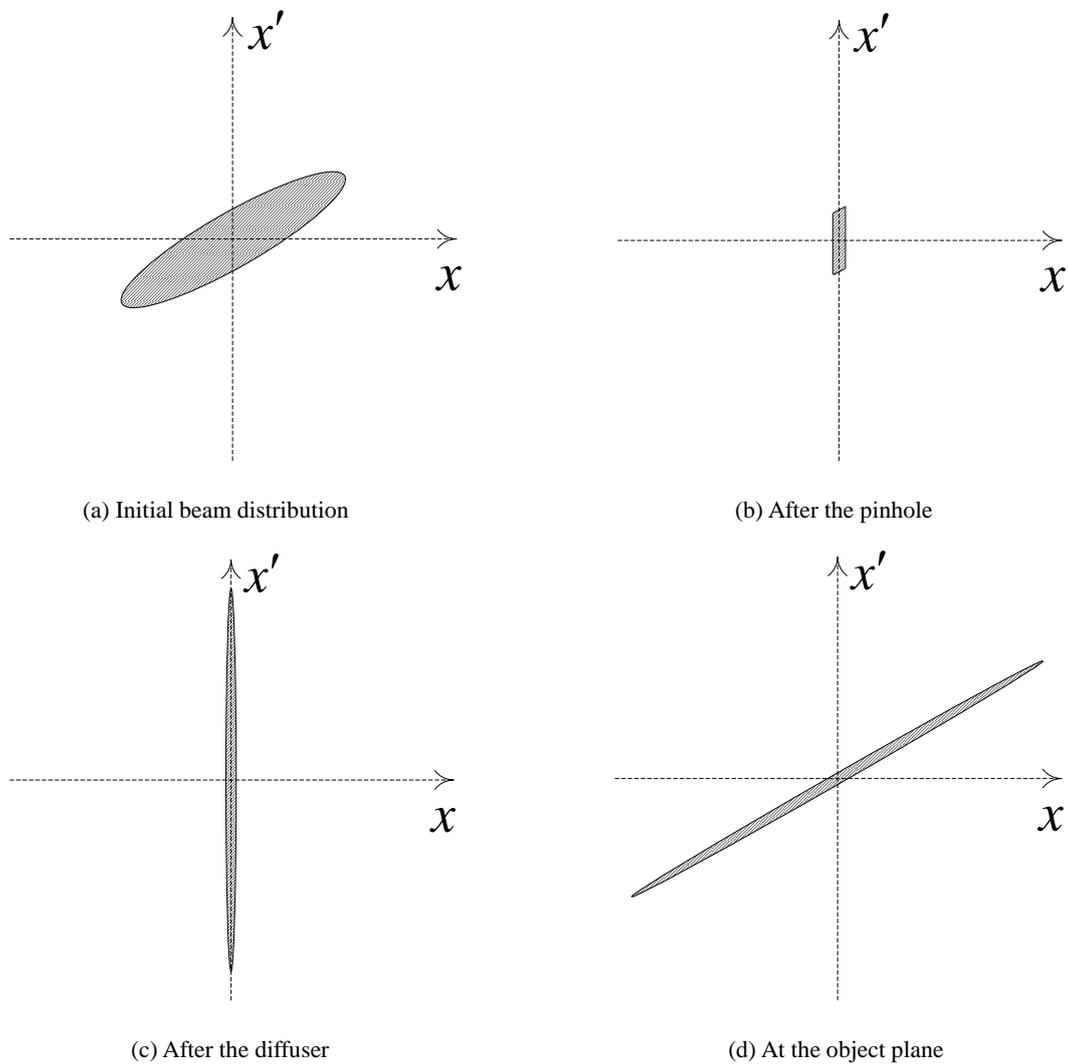

(a) Initial beam distribution  (b) After the pinhole

(c) After the diffuser  (d) At the object plane

Fig.4　Evolution diagram of transverse phase space. (a) is the initial horizontal phase space, (b) is the beam distribution after the pinhole, (c) is the case after the diffuser and (d) is the case at the object plane.

The Zumbro lens system consists mainly of four identical quadrupole magnets, after proper arranging, it can achieve one-to-one imaging. The aperture is 60mm and the chromatic coefficients [7] $R'_{12}$ (for x direction) and $R'_{34}$ (for y direction) are both about 2.8m. The collimator adopts hole-type structure, the inner hole is empty and proton beam can travel through, the outer layer is made of aluminum which can prevent protons. With the collimator, one can limit the transmitted particles to only those with a MCS angle less than the cut angle.

The imaging system is made of a piece of LSO scintillator and followed CCD camera. Visible light is generated when the proton beam hitting the scintillator and collected by a CCD camera.

3、Radiographic performance

The low energy pRad system was commissioned in June 2013 at Institute of Fluid Physics. The imaging beamline performed as designed. Detailed measurement of spatial resolution was collected by studying the pRad image of zebra crossing as shown in figure 5. The stripes with different widths were printed on aluminum foil by laser printer, and the partition length is equal to the width of each stripe. There are four kinds of stripes in figure 5, the widths are 1mm, 0.5mm, 0.2mm and 0.1mm separately. In order to obtain clear image, the beam distribution should be uniform in x-y plane, but it is hard to achieve. In fact, we can remove the influence of beam's uneven distribution via the way original radiograph divided by the beam radiograph. Moreover, the noise introduced by CCD should be removed also. As shown in figure 5, the spatial resolution of 11MeV low energy pRad is better than 0.1mm.

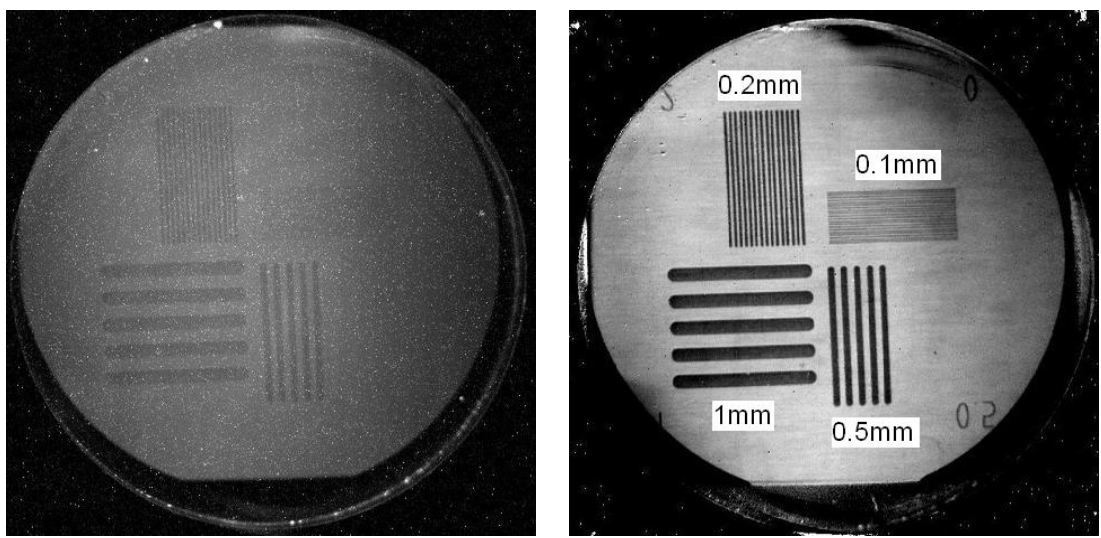

Fig.5    Low energy pRad image of the stripes on aluminum foil. The left one is the original radiograph, the right one is the image which removes the influence of noise and beam's uneven distribution.

In addition to detailed resolution measurements some familiar items were radiographed as a demonstration of the sensitivity and high resolution capabilities of the radiography system. A radiograph of cicada wings is shown in figure 6, The significant contrast shown in this radiograph demonstrates the extreme sensitivity of low energy pRad system to objects with thin areal densities.

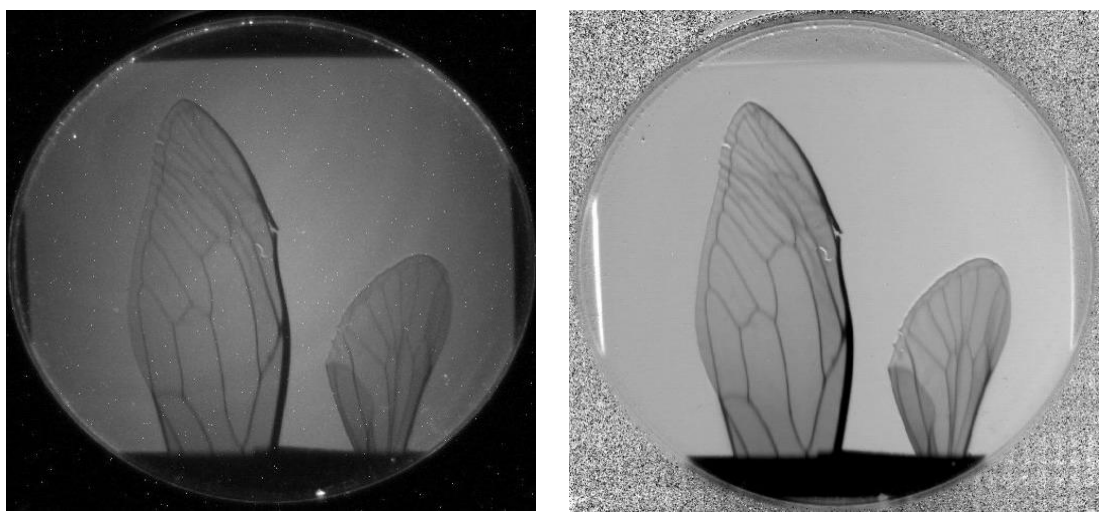

Fig.6    Low energy pRad image of the cicada wings. The left one is the original radiograph, the right one is the image which removes the influence of noise and beam's uneven distribution.

## 4、Conclusion

The 11MeV low energy pRad system can provide scatheless diagnosis for static objects with 0.001~0.03g/cm$^2$ areal densities with spatial resolution of ~100μm. Future upgrades to this system could improve the spatial resolution remarkably by the using of magnifying Zumbro lens system [8].